\documentclass[]{./aa}

\usepackage[final]{epsfig} 

\usepackage{times}

\usepackage{natbib}
\bibpunct{(}{)}{;}{}{}{,}

\usepackage[american,german,english]{babel}

\usepackage{latexsym}
\usepackage{amssymb}

\begin{document}
\title{Extending the limits of globule detection%
  \thanks{Based on observations with ISO, an ESA project with instruments
    funded by ESA Member States (especially the PI countries: France,
    Germany, the Netherlands and the United Kingdom) and
    with the participation of ISAS and NASA.}
      }

\subtitle{ISOPHOT Serendipity Survey observations of interstellar clouds}
\author{L.V.~T\'oth\inst{1,2,3}
  \and  Cs.~Kiss\inst{4}
  \and  M.~Juvela\inst{1}
  \and  M.~Stickel\inst{3} 
  \and  U.~Lisenfeld\inst{5}  
  \and  S.~Hotzel\inst{3} 
}

\institute{
  Helsinki University Observatory, P.O.\,Box~14, 
  FIN-00014 University of Helsinki, Finland
  \and 
  Department of Astronomy, Lor\'and E\"otv\"os University,
  P\'azm\'any P\'eter s\'et\'any~1/a, H-1117~Budapest, Hungary
  \and
  Max-Planck-Institut f\"ur Astronomie, K\"onigstuhl~17,
  D-69117~Heidelberg, Germany
  \and   
  Konkoly Observatory, Budapest, PO Box 67, H-1525, Hungary  
  \and 
  IRAM, Avenida Divina Pastora 7, N.C., 18012 Granada, Spain
  }
\offprints{L.V..~T\'oth e-mail:lvtoth@mpia-hd.mpg.de}

\date{Received Feb 4, 2002; accepted ...}

\abstract{
A faint $I_{\rm 170}=4$\,MJysr$^{-1}$ bipolar globule was discovered with
the ISOPHOT 170\,$\mu $m Serendipity Survey (ISOSS).
ISOSS\,J\,20246+6541 is a cold ($T_{\rm d}\approx 14.5$\,K) FIR source 
without an IRAS pointsource counterpart.
In the Digitized Sky Survey B band it is seen as a 3\arcmin size bipolar nebulosity with an average excess surface 
brightness of $\approx 26$\,mag/$\square $\arcsec . 
The CO column density distribution determined by multi-isotopic, 
multi-level CO measurements with the IRAM-30m telescope
agrees well with the optical appearance. 
An average hydrogen column density of $\approx 10^{21}$cm$^{-2}$ 
was derived from both the FIR and CO data.
Using a kinematic distance estimate of 400\,pc
the NLTE modelling of the CO, HCO$^+$, and CS measurements gives a 
peak density of $\approx 10^4$cm$^{-3}$.
The multiwavelength data characterise 
ISOSS\,20246+6541 as a representative of a class of globules which has not 
been discovered so far due to their small angular size and low 100$\mu $m 
brightness.
A significant overabundance of $^{13}$CO is found
$X(^{13}$CO$) \ge 150\times X($C$^{18}$O$)$. This is likely due to isotope 
selective chemical processes.
}
\maketitle
%
\section{Introduction\label{introduction}}
Bok globules were originally detected in absorption against HII regions by 
\citet{bok}, and are known as small, dense interstellar clouds in 
the solar neighbourhod ($d\lessapprox 400$pc). They were identified
optically, and mostly the nearby ones have been catalogued so far. 
It is expected that they are similarily common elsewhere in
the Galactic disk and in fact a few distant ones are already investigated eg. 
by \citet{launhardt}. 
Their FIR properties were determined by \citet{clemens} based on IRAS
data. They emphasized the importance of finding FIR faint cold globules
since these are representatives of the inactive (i.e. non-starforming) 
interstellar medium. Starless globules with small apparent size and
low temperature (thus also very low 100\,$\mu $m brightness) can be seen
only by good sensitivity at wavelengths over 100\,$\mu $m.
An ISOPHOT study of pre-stellar cores in dark clouds with low 
100\,$\mu $m brightness were recently reported by \citet{ward}.
The ISOPHOT 170\,$\mu $m Serendipity Survey (ISOSS)
can be used to locate cold galactic objects even without 
any preliminary identification. 
This raises the possibility of detection of the "missing" globules via ISOSS.
We present our results on ISOSS\,J\,20246+6541 proving that it 
is, indeed, a small and cold isolated 
molecular cloud - one of the so far missed population.

\section{Observation and data analysis}

We searched ISOSS at medium galactic latitudes for 170\,$\mu$m 
point sources without IRAS counterparts. The ISOSS measurements,
calibration, and data analysis for interstellar clouds are 
described by \citet{bogun_96}, \citet{stickel_00} and \citet{toth_00}
respectively.
ISOSS\,J\,20246+6541 was detected as an 
$I_{170}(\rm peak)= 4$\,MJysr$^{-1}$ ($>4\sigma $) pointlike 
(FWHM $\le 2.5$\arcmin) 
source at RA(2000)=20$^{\rm h}$24$^{\rm m}$36$^{\rm s}$ 
Dec(2000)=+65$\degr$40$\arcmin$ (l=99.80$\degr$, b=15.70$\degr$).
It is without an IRAS point source counterpart.

The optical appearance of CISS1 and its neighourhood was studied using
40\arcmin\,$\times $\,40\arcmin\ Digital Sky Survey (DSS) blue and red images.
Photometry on DSS plates of all USNO \citep{usno} sources have 
been done where 
R and B $<$ 18mag. 
Integrated photographic density was derived for all these USNO sources. 
Photographic density values DN of the background were
measured at each plate at 25 positions where neither star, nor 
significant background emission enhancement was seen.
The background value is $DN(BG)_R =4590\pm 190$ and $DN(BG)_B=4610\pm 230$
in photographic density units for R and B bands respectively.
Photometric calibration of the plates was made following Cutri's (1993)
method. The photographic density to magnitude calibration formulae 
were derived as $log_{10}({DN})$ vs. magnitude relations.
The scatter is large for faint stars, and the
extrapolation to low photographical density (i.e. high magnitude) 
values results in a large error bar. 
In order to make surface brightness maps of ISOSS\,J\,20246+6541, 
stars were removed from the optical images by substituting 
the average surrounding DN value. 

\begin{figure}
\epsfig{file=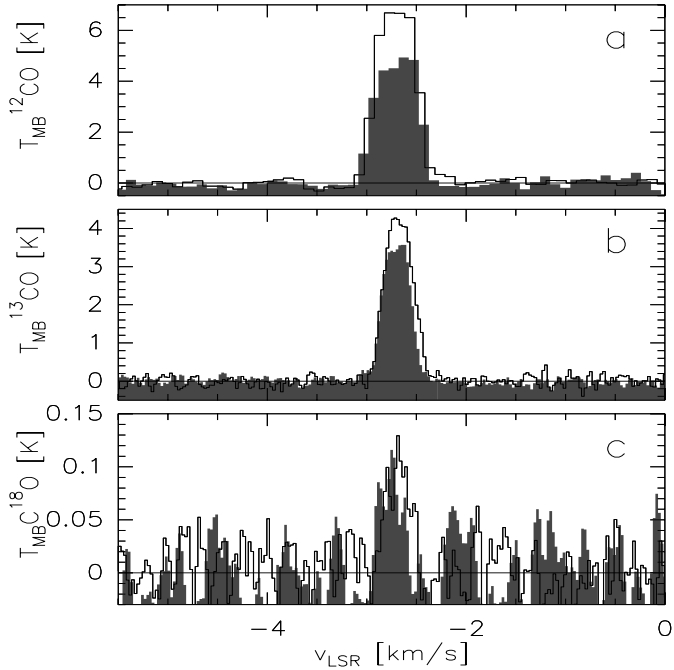, width=\hsize, height=\hsize}
\caption{CO spectra at the $^{13}$CO(1--0) peak, (2--1) lines are 
overlaid as filled histograms.
         {\bf a,} $^{12}$CO(1--0) and (2--1) spectra.
         {\bf b,} $^{13}$CO(1--0) and (2--1) spectra.
         {\bf c,} C$^{18}$O(1--0) and (2--1) spectra.
  \label{cospectra}}
\end{figure}

The $J=$(1--0) and (2--1) rotational lines of CO, $^{13}$CO and C$^{18}$O
were measured in Aug.~1998 with the IRAM-30 telescope.
A $4.5\arcmin \times 4.5\arcmin$ region centered on ISOSS\,20246+6541
was mappen in on-the-fly mode 
in $^{12}$CO(1--0), $^{12}$CO(2--1) and $^{13}$CO(1--0) 
with angular resolutions
of 20\arcsec , 10\arcsec, 20\arcsec\  respectively.
Pointed measurements
were performed in all the above mentioned transitions at the
$^{13}$CO(1--0) peak position 
(22$^{\rm h}$24$^{\rm m}$46$^{\rm s}$, 65\degr40\arcmin01\arcsec). 
Spectral resolutions of 0.1, 0.05 and 0.03kms$^{-1}$ were used.
These spectra are shown in 
Fig.~\ref{cospectra}.
The lines are narrow with a FWHM of $\le 0.5$kms$^{-1}$.
The result of the 1998 C$^{18}$O pointed measurements  
were confirmed in Sept. 2000 when also the 
CS(2--1) and (3--2) as well as the HCO$^+$(1--0) and (2--1) transitions
were observed with the IRAM-30 telescope. The C$^{18}$O, CS(2--1) and 
HCO$^+$(1--0) lines were detected with $S/N>5$ and well resolved with
the 0.03kms$^{-1}$ spectral channels.
The data were calibrated to the scale of corrected antenna temperature,
$T^*_{\rm A}$, by observing loads at ambient and cold temperature, as in the 
conventional ``chopper-wheel'' calibration for millimetre wavelength
observations. 
The data were converted to the main-beam scale by
applying the relation $T_{\rm MB} = 
T^*_{\rm A}\times (B_{\rm eff}/F_{\rm eff})$.
The values for the main-beam efficiency, $B_{\rm eff}$,  
are 0.70 and 0.42 (July 1998), and 0.75 and 0.53 (September 2000), 
and the values for the forward efficiency,
$F_{\rm eff}$, are
0.92 and 0.85 for CO(1--0) and CO(2--1), respectively.
We checked the calibration by observing the standard sources
DR~21 and W3OH and found it always to be better than 20 \%.

\section{Results\label{results}}

\subsection{FIR results and derived parameters}

ISOSS\,J\,20246+6541 is a faint FIR source with an upper limit on 
the $I_{100}$ brightness of about 0.5\,MJysr$^{-1}$  derived from 
the ``raw'' (IRDS format) IRAS data. 
Comparison of ISOSS and ISSA data \citep{wheelock_94} was made
as described by \citet{toth_00}. An upper limit of the colour 
temperature of 14.5\,K was estimated from the bisector slope of the
 $I_{170}$ vs. $I_{100}$ scatter plot.
Assuming a dust temperature of 14.5\,K, an average
dust column density of $1.3\times 10^{-5}$gcm$^{-2}$ was derived
following \citet{hildebrand_83}.
This corresponds to an average hydrogen column density of 
$N($HI+2H$_2)=8.6\times 10^{20}$cm$^{-2}$ where a 
hydrogen-to-dust mass ratio of 110 \citep{launhardt} was assumed.

\subsection{Optical images}
ISOSS\,J\,20246+6541 appears as a faint, isolated reflection cloud west 
of the L1155/L1157 cloud complex. When smoothed to 15\arcsec\ 
the excess diffuse surface brightness distribution  of ISOSS\,20246+6541
shows a ``bright'' lobe at the NE and a fainter 
fragmented one at SE. The POSS B band diffuse surface brightness is
shown in Fig.~\ref{surfaceb}\,a, where the contours are drawn from 
22.7\,mag/${\square }\arcsec$ by $-0.05$mag/${\square }\arcsec$ .
The lowest contour corresponds to three times the
standard deviation of photographical density outside the globule 
on the star-removed, smoothed image.
\begin{figure*}
\epsfig{file=CISS1maps.epsi, height=6cm}
\caption[]{Dust and gas in ISOSS\,20246+6541, 
reference coordinate (0,0): 
RA=$20^{\rm h}24^{\rm m}44^{\rm s}$ DEC=$65\degr 40\arcmin04\arcsec$\\
{\bf a:} B band diffuse surface brightness, contours are from 
22.7\,mag$/\square \arcsec$ (i.e. background level + $3\times \sigma $ 
converted to magnitudes) 
 $-0.05$mag$/\square \arcsec$\\
The diffuse surface brightness contours indicate a sharp cloud edge at NE.
The structures at offset coordinates (-0.5, -1.8) are less reliable
due to residuals of a removed 11.9\,mag star GSC\,04245-00803.
{\bf b:} $W(^{12}$CO\,(1--0)), contours are from 1.5\,Kkms$^{-1}$ with
        1.5\,Kkms$^{-1}$ steps; 
{\bf c:}$^{13}$CO\,(1--0), contours are from  1.0\,Kkms$^{-1}$ with
        1.0\,Kkms$^{-1}$ steps;
CO isotopomer line intensities are integrated in the velocity interval 
-5\,kms$^{-1}$\,$\le$\,v$\rm _{LSR}$\,$\le$\,0\,kms$^{-1}$.
\label{surfaceb}}
\end{figure*}
The average background surface brightness is 22.8/${\square }\arcsec$.
Linear interpolation of the \citet{leinert} table of the B band
sky brightness distribution gives  
at the the cloud position 102\,$S_{10}$, which agrees well with our result.
The average excess surface brightness of ISOSS\,J\,20246+6541
is $\approx 26$\,mag/$\square $\arcsec in the DSS B band. 
The same morphology is shown by  
the B and R surface brightness distributions, except that the
SE lobe is slightly more red than the NW lobe. 
The lack of a dark core indicates an average visual
extinction below $1.5$\,mag.

\subsection{Molecular line results, and derived parameters}

All observed lines show a LSR velocity of -2.7kms$^{-1}$. The 
FWHM widths of the $^{12}$CO lines are around 0.4kms$^{-1}$, 
and around 0.3kms$^{-1}$ for all the other detected lines.
The $^{12}$CO and $^{13}$CO integrated intensity distributions are shown in 
Figs.~\ref{surfaceb} b and c.
The bipolar shape is well seen and the CO line intensities are in
accordance with the excess surface brightness distribution,
the NE lobe being much brighter. The two lobes show the same radial
velocity.

The physical parameters were at first approximated 
from the $^{12}{\rm CO}(1-0)$ and $^{13}{\rm CO}(1-0)$ spectra of the
peak intensity positions,
assuming an isothermal spherical cloud in LTE which uniformly fills the beam.
The LTE CO(1-0) excitation temperature is $T_{\rm ex}($CO$)=9.2$\,K,
characteristic of non-starforming molecular clouds.
We derived an optical depth of $\tau (^{13}{\rm CO})=0.87$ from the observed 
CO(1-0) and $^{13}$CO(1-0) line ratio assuming the terrestrial 
isotopic density ratio of 89.
The CO column density was calculated using the standard LTE formula:\\
$N(^{13}{\rm CO})=2.57\times 10^{14}T_{\rm kin}
\frac{\tau (^{13}{\rm CO}) \Delta v}{1-\exp (-\frac{h\nu }{kT_{\rm kin}})}$,
resulting in
$N(^{13}{\rm CO})=8.9\times 10^{14}$cm$^{-2}$,
which corresponds to
$N({\rm H}_2)\approx  10^{21}$cm$^{-2}$ assuming
$N({\rm H}_2)\approx 10^6 N^{13}({\rm CO})$.
This result is in agreement with the column density derived
from the FIR data.
The $^{12}$CO and $^{13}$CO lines trace the ISM well over
most of the cloud.
A comparison of C$^{18}$O to $^{13}$CO lines (both 1-0 and 2-1)
at the centre of the NE lobe indicates an underabundance of 
C$^{18}$O by a factor of 4. This effect is expected in cold clouds 
with moderate density,
exposed to UV radiation \citep{glassgold}. A more careful modelling
may account for it as we show in the Discussion session.

\section{Discussion}

\subsection{Distance of ISOSS\,J\,20246+6541}
The distance of ISOSS\,20246+6541 can be estimated, relating it 
to its neighbours. Its nearest neighbours are \object{L1122} and 
\citep{lynds_62}, and the \object{YDM97~CO1} \citep{yonekura_97}.
The $^{13}$CO survey of Cepheus by \citet{yonekura_97} covers the 
position of ISOSS\,J\,20246+6541 and 
their Fig.~6\,a indicates few small clouds around ISOSS\,J\,20246+6541
(i.e. \object{YDM97~CO1}, \object{YDM97~CO2}, \object{YDM97~CO3})
 and one even smaller  unnumbered peak 
very close to ISOSS\,J\,20246+6541 at l$=100\fdg1$ b$=15\fdg7$.
All the listed Yonekura clouds                 
have $v_{\rm LSR}>+2.5$kms$^{-1}$, and they are counted into the 
``close group'' of clouds, which on the other hand is associated
with  extended FIR features around ISOSS\,J\,20246+6541.
The nearest molecular clouds with negative $v_{\rm LSR}$ are
 \object{YDM97~CO7, CO9, CO10} at $l\approx 103$\fdg0 $b\approx 16$\fdg7.
ISOSS\,20246+6541 itself has $v_{\rm LSR}=-2.7$kms$^{-1}$.
It probably belongs to the ISM layer of the nearby Cepheus Flare GMC, 
and is located at about 400pc \citep{kun_98}. 
We note that applying the size-linewidth relation of
\citet{larson} the globule may be between 100 and 400pc.

\subsection{Radiative transfer models}

We have modelled the NE lobe of the bipolar globule with spherically
symmetric cloud models, although the NE clump shows some
deviations from spherical symmetry in both $^{12}$CO and $^{13}$CO
(see Fig.~\ref{surfaceb} b and c.
With RA=20$^{\rm h}$24$^{\rm m}$44$^{\rm s}$ 
Dec=+65$\degr$40$\arcmin$04$\arcsec$ as the
centre position, we have averaged spectra in concentric rings
with radii increasing by 10$\arcsec$ intervals up to a radii of 90$\arcsec$. 
The effective resolution of the averaged spectra
 is 40$\arcsec$ for the $J=$1--0
lines and 20$\arcsec$ for the $J=$2--1 lines.

We set the cloud parameters as follows.\\
(1) We assume a density distribution $n \sim r^{-1.5}$ with a density ratio
100 between the centre and the cloud surface. \\
(2) The kinetic temperature is assumed to rise linearly from the cloud
centre. This is a crude
approximation of the actual temperature structure of for a
small, spherically symmetric globule without internal heating sources
(e.g. Leung 1985; Nelson \& Langer 1999) 
but will suffice for the present purposes.
The temperature gradient, i.e. the difference between the 
outermost and innermost shells 
$\Delta T=0$\,K, 6\,K or 10K. Higher contrast 
means too high a temperature for the outer cloud, in contradiction with 
the observed small linewidth.\\
(3) Extinction-dependent relative molecular abundances 
$X({\rm molecule})=\frac{n({\rm molecule})}{n(H)}$ were 
estimated according to \cite{warin}. 
The cloud is cold, exposed to UV radiation and it has a peak 
visual extinction between 1 and 2 mag. 
In these conditions isotope selective processes
result in a relative overabundance of $^{13}$CO and relative 
underabundance of C$^{18}$O according to \cite{bally}.
When applying the \cite{warin} relative abundances, we
introduced an intrinsic extinction at the cloud boundary since the
$^{12}$CO lines are not vanishing at the boundary of the NE lobe.
This assumption is supported by the presence of surrounding
extended cirrus-like emission seen in 100\,$\mu $m on the ISSA image.\\
(4). Distance:  100, 200, 400, 600, 800, 1000, and 2000pc
were tested.\\

\begin{figure}[h]
\epsfig{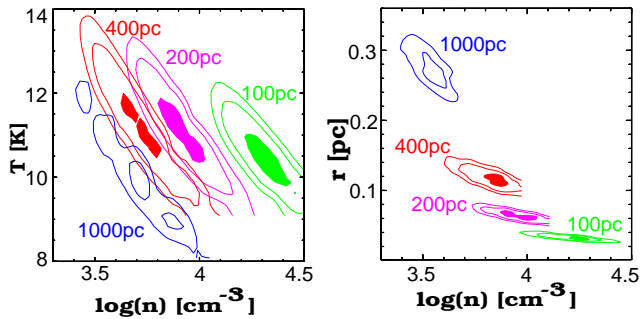}

\caption{Results of the radiative transfer modelling of the northern lobe
 assuming distances: 100pc (yellow), 200pc (lilac), 400pc (red), 
1000pc (blue). The relative $\chi ^2$ minimum regions are plotted
as shaded regions in selected planes of the parameter space.
Two further contours of the relative $\chi ^2$ of 1.14 \& 1.29 are overlaid.
(Extinction dependent relative molecular 
abundances were used.)
{\bf a:} density versus kinetic temperature (left); 
{\bf b:}density versus size (right)
\label{model}}
\end{figure}

When the density, temperature, relative abundance distributions and 
the distance is set to a value allowed by the above constrains, the
free model parameters are the central density ($n_{\rm c}$), the
intrinsic linewidth ($\Delta v$) and the angular diameter ($D$) of
the model cloud.
The radiative transfer problem is solved with Monte Carlo simulation
\citep{juvela}. The computed spectra are convolved to the
resolution of the observed spectra and the quality of the fit between
the two is estimated with a weighted $\chi^2$ value. 
The model cloud is divided into 31 shells of equal thickness and the
free parameters are optimized separately for $^{12}$CO and $^{13}$CO.
We then select the set of parameters which provides the best fits for both.
Since the $^{12}$CO observations only probe the outer layers
of the cloud, the $^{12}$CO based estimate of the column density 
is uncertain. 
Modelling based on the $^{13}$CO line,
however, gives surprisingly similar results when the appropriate average
relative abundance value $X(^{13}$CO$)\approx 1.1\times 10^{-6}$ is selected.
The models are not sensitive to 20\% changes in the average molecular
abundances or density, although similar changes of the kinetic 
temperature or size are, critical (see Fig.~\ref{model}).

Synthetic spectra for C$^{18}$O CS and HCO$^+$ were generated
with the NLTE model using the best parameter sets (lowest $\chi ^2$) 
from the $^{12}$CO $^{13}$CO analysis. The relative abundances were
varied up to 100\% and the other parameters up to 30\%.
The C$^{18}$O lines were best reproduced
assuming an average relative abundance of 
$X(^{13}{\rm CO})/X({\rm C}^{18}{\rm O})\approx 150$, an extreme but
possible underabundance by factor of 28 \citep{glassgold}.
The pointed
measurements supported the density and temperature results shown in
Fig.~\ref{model}.
 The derived NLTE kinetic temperature is around 11K and assuming a 
distance of 400\,pc 
the peak hydrogen density and the size of the NE lobe is
$n_{\rm c}$=6.7$\cdot$10$^4$\,cm$^{-3}$ and 0.12\,pc respectively. 
The column density estimate is 
$N({\rm H}_2)\approx 2\cdot 10^{21}$\,cm$^{-2}$. The total gas
mass would be $\approx 3{\rm M}_{\odot }$.

With $N({\rm H}_2) \sim$10$^{21}$\,cm$^{-2}$,  the visual extinction
towards the cloud centre is $A_{V} \approx$1 \citep{bohlin} 
and for UV photons the cloud is optically thick unless it is very clumpy.  
External heating, however, is reduced by the
surrounding ISM, which is represented by the nonvanishing $^{12}$CO 
lines. This may be the reason that a moderate 6K temperature contrast was
found to be more likely than a 10K contrast or isothermal.

Although the SW clump is clearly elongated similar modelling was
carried out for that part of the cloud. Observed spectra were
averaged over annuli at radii up to 50$\arcsec$ from the clump
centre. Assuming a model where the kinetic temperature increases
linearly from the centre, we obtain a peak column density of 
$N({\rm H}_2)$=2.9$\cdot$10$^{20}$\,cm$^{-2}$ based on the CO
spectra.

\section{Concluding remarks}
\begin{enumerate}
\item ISOSS\,20246+6541 is a small isolated molecular cloud discovered by
its 170\,$\mu $m emission.
\item The globule appears as bipolar in both optical surface brightness and
mm-line maps.
\item A distance of 400 pc is likely, setting the diameter of the 
northern lobe to approximately 0.1 pc.
\item The average physical parameters are: 
$T_{\rm kin}=11$K, $n($H$_2)=7000$cm$^{-3}$.
\item In the $5\degr $ environment of ISOSS\,20246+6541 
we found further 15 ISOSS sources with similar FIR parameters.
One of those is another previously unknown globule as seen on DSS2 images.
\item The population of small and faint starless globules can only 
be explored by high sensitivity FIR measurements such as those carried out by
PHT-C2 on board ISO.
\item We note that there are no other ISO measurements but the 
ISOSS within 40\arcmin
distance from ISOSS\,J\,20246+6541. Our finding thus also demonstrates
the importance of such ubiased surveys by future missions like Planck.
\end{enumerate}

\begin{acknowledgements}
\sloppy
We acknowledge the numerous valuable comments by Prof. Kalevi Mattila,
and important notes by Dr. Mark Rawlings.
The ISOPHOT project and Postoperation Phase
was funded by the Deut\-sche
Agen\-tur f\"ur Raum\-fahrt\-an\-ge\-le\-gen\-heiten (DARA, now DLR), 
the Max-Planck-Gesell\-schaft, the Danish, British and Spanish Space
Agencies and several European and American institutes.\\
Members of
the Consortium on the ISOPHOT Serendipity Survey (CISS) are MPIA
Heidelberg, ESA ISO SOC Villafranca, AIP Potsdam, IPAC Pasadena,
Imperial College, London.\\
L.\,V.~T\'oth acknowledges an MPI research fellow grant.
This research was partly supported by the OTKA F-022566 grant
and by the Academy of Finland through grants No.\ 158300 and
173727. \\
This research has made use of the
Digitized Sky Survey, produced at the Space Telescope Science
Institute, NASA's Astrophysics Data System Abstract Service, the
Simbad Database, operated at CDS, Strasbourg, France.\\
\end{acknowledgements}



\end{document}